\documentclass{article}

\usepackage[dvips]{graphicx}

\usepackage{amssymb,amsfonts,amsmath}

\def\reactionrates#1{\mathrel{\mathop{\rightarrow}\limits^{#1}}}

\begin{document}
\begin{titlepage}
\begin{center}
\Large{Mapping between dissipative and Hamiltonian systems}

\normalsize

 Jianhua Xing  \\
Department of biological sciences, Virginia Polytechnic Institute and State University, Blacksburg, VA 24061

\end{center}


\begin{abstract} 
Theoretical studies of  nonequilibrium systems are complicated by the lack of a general framework. In this work we first show that a transformation introduced by Ao recently (J. Phys. A {\bf 37}, L25 (2004)) is related to previous works of Graham (Z. Physik B {\bf 26}, 397 (1977)) and Eyink {\it et al.} (J. Stat. Phys. {\bf 83}, 385 (1996)), which can also be viewed as the generalized application of the Helmholtz theorem in vector calculus. We then show that systems described by ordinary stochastic differential equations with white noise can be mapped to thermostated Hamiltonian systems. A steady-state of a dissipative system corresponds to the equilibrium state of the corresponding Hamiltonian system. These results provides a solid theoretical ground for corresponding studies on nonequilibrium dynamics, especially on nonequilibrium steady state. The mapping permits the application of established techniques and results for Hamiltonian systems to dissipative non-Hamiltonian systems,  those for thermodynamic equilibrium states  to nonequilibrium steady states. We discuss several implications of the present work. 
\end{abstract}
\end{titlepage}

\clearpage

\clearpage
\section{Introduction}
Equilibrium statistical mechanics is a fundamental branch of physics with applications extending widely through the scientific domain. Based on several fundamental laws or hypothesis, the theoretical frames are self-consistent and complete.  
 However, most systems in nature are not at equilibrium. Open systems, which exhibit free energy dissipation and thus no detailed balance, are of special interest. Our understanding of nonequilibrium systems is still primitive. 
 Due to the lack of detailed balance, the methods used to study nonequilibrium processes lack general or systematic approaches
and often depend on the specific context.  Several attempts have been made to develop a unified framework and establish a connection between equilibrium and nonequilibrium phenomena \cite{ Fitts1962,Glansdorff1971, Landsberg1972,Landauer1973,Jou1993,deGroot1984,Keizer1987,Seifert2008}. 
 For example, Oono and Paniconi noticed the similarity between an thermodynamic equilibrium state and a nonequilbrium steady state, and proposed an operationally feasible phenomenological framework called "steady-state thermodynamics" \cite{Oono1998}.  Hatano and Sasa applied the formalism to Langevin systems. They and others derived several useful results  \cite{Hatano2001,Trepagnier2004,Prost2009}. 
 
 Continuous efforts have been made but with limited success  to map a non-Hamiltonian  to a Hamiltonian dynamical system \cite{ Morrison1986,Volterra1931, Martin1973,Tailleur2008}. As Morrison pointed out, one important reason for the research is that knowing the Hamiltonian structure  of a complex system provides guidance for studying the dynamics \cite{Morrison1998}.
 
A related topic it is highly debated issue of the existence of potential functions for dynamic systems without detailed balance \cite{Graham1971, Grmela1997} . On the other hand, the metaphor of energy landscapes has long be adopted to understand evolution of some biological systems \cite{Wright1932,Waddington1940,Delbruck1949, MacArthur2009, Huang2009}.  Recently, Ao has proposed a useful way of decomposing a set of stochastic differential equations (SDEs) and construct dynamic potentials \cite{AoJPhys2004}.   Ao and coworkers examined explicitly some linear systems for which analytical solutions can be obtained \cite{Kwon2005}. However, the validity of his procedure for nonlinear systems is obscured by the abstract presentation. 
 
 For clarity, we  we now summarize the content and the main results of this work. First, we will  demonstrate that for systems possessing steady states, one can follow some standard procedure to decompose the effective flux of the corresponding Fokker-Planck equations\cite{Graham1977, Eyink1996}.  Ao's procedure turns out  to be equivalent to a special choice of the effective flux. It is essentially a Helmhotz-Hodge decomposition for the corresponding stochastic differential equations. The conclusion is {\it not} restricted to linear systems. 
Next we will prove that there exists  a mapping between stochastic dissipative systems and Hamiltonian systems, nonequilibrium steady-state and thermodynamic equilibrium state, respectively. To our knowledge this is the first rigorous and general proof of the mapping. 
The mapping reveals common mathematical structures of classes of dynamic systems with and without detailed balance. It serves as the theoretical basis for 1) the phenomenological steady-state thermodynamics formulations parallel to equilibrium systems; 2) the simplified derivation of many corresponding relations following standard procedures discussed in any introductory equilibrium statistical physics textbooks;  3) the concept of potential landscapes which has been fruitfully applied in biological applications.  Unlike that of Oono and Paniconi, our discussion is {\it not} restricted to steady states and their vicinity, but applies to dynamics arbitrarily far from the steady states. Then we will present specific examples, and conclude the paper with discussions on the implications of this work and potential future studies. 
 
 \section{Theory}
Similar to Hatano and Sasa,  we will consider processes described by the following form of SDEs,
\begin{eqnarray}
dx_i/dt = G_i(\mathbf{x})+\epsilon \sum_{j=1}^m g_{ij}(\mathbf{x}) \zeta_j(t), i=1,\cdots, n. \label{eqn:stochaseqn}
\end{eqnarray}
In general $m$ and $n$ may be different,  $\zeta_i(t)$ are temporally uncorrelated, statistically independent Gaussian white noise with the averages satisfying $<\zeta_i (t) \zeta_j(\tau)> =  \delta_{ij} \delta(t-\tau)$, 
$\mathbf{g}(\mathbf{x})$ is related to the $n\times n$ diffusion matrix $\mathbf{gg}^T=2 \mathbf{D}$, where the superscript $T$ refers to transpose. For a physical system $\epsilon$ is the Boltzmann's constant multiplying temperature. 
For a non-physical system, one can define an effective temperature relating to $\epsilon$.  
 To help the general readers, let's consider an example reaction series, 
$\reactionrates{k_0}A\reactionrates{k_1}B\reactionrates{k_2}$, the chemical Langevin equations are $d[A]/dt = k_0-k_1[A]+\sqrt{k_0}\zeta_{1}/\sqrt{\Omega} - \sqrt{k_1[A]}\zeta_{2}/\sqrt{\Omega}$, 
  $d[B]/dt = k_1[A]-k_2[B]+\sqrt{k_1[A]}\zeta_{2}/\sqrt{\Omega} - \sqrt{k_2[B]}\zeta_{3}/\sqrt{\Omega}$, where $\Omega$ is the system volume \cite{vanKampen2007, Gillespie2000}.
Equations in the form of Eqn. \ref{eqn:stochaseqn} have been applied to problems from physics, chemistry, cellular biology, ecology, engineering, finance, and many other fields \cite{vanKampen2007, Gardiner2004, Hanggi1990,Cobb1981}.  

In a recent paper  \cite{AoJPhys2004}, Ao argued that one can always construct a symmetric matrix $\mathbf{S}$ and an anti-symmetric one $\mathbf{T}$, and transform Eqn. \ref{eqn:stochaseqn} into,
\begin{eqnarray}
(\mathbf{S}+\mathbf{T})\cdot\frac{d\mathbf{x}}{dt}  &=& (\mathbf{S}+\mathbf{T})\cdot (\mathbf{G(x)} +\epsilon \mathbf{g(x)}\cdot\zeta(t) ) \nonumber\\
&=&  - \nabla_\mathbf{x} \phi(\mathbf{x}) +\epsilon \mathbf{g}'(\mathbf{x})\cdot\zeta(t) \label{eqn:transeqn}
\end{eqnarray}
where $\phi$ is a scalar function corresponding to the potential function in a Hamiltonian system satisfying 
$(\partial \times \partial\phi)_{ij}\equiv (\partial_i\partial_j-\partial_j\partial_i)\phi=0$. To uniquely  determine the 
transformation matrices $\mathbf{S}$ and $\mathbf{T}$, Ao introduced additional constraints 
$\mathbf{g' g'}^T = 2\mathbf{S}$. This is similar to choosing a gauge in electrodynamics (see below). Then $\mathbf{S}$ and $\mathbf{T}$ are determined by 
\begin{eqnarray}
\partial\times[\mathbf{(M\cdot G(x)}]=0,(\mathbf{M})^{-1}+(\mathbf{M})^{-T}=2\mathbf{D}, \label{eqn:M_eqn}
\end{eqnarray}
where $\mathbf{M=S+T}$. The above relation also suggests  $\mathbf{M^{-1}=D+Q}$, where $\mathbf{Q}$ is antisymmetric. One can identify the symmetric and anti-symmetric matrices as the friction and transverse matrices, respectively \cite{AoJPhys2004}.  This transformation has be applied to network studies \cite{Zhu2004, Wang2006}.
On can further introduce the inertial term  \cite{AoJPhys2004}, and transform Eqn. \ref{eqn:transeqn} into,
\begin{eqnarray}
\frac{d\mathbf{x}}{dt}  &=& \frac{\mathbf{p}} {m}, \label{eqn:xdot}\\
 \frac{d\mathbf{p}}{dt}  
  &=& \left[-\mathbf{T(x)} \cdot \frac{\mathbf{p}}{m} -\nabla_\mathbf{x}\phi(\mathbf{x})\right] \nonumber\\ 
 &&  + \left[- \mathbf{S(x)}\cdot \frac{\mathbf{p}}{m} +\epsilon  \mathbf{g}'(\mathbf{x}) \cdot \zeta(t) \right]  \label{eqn:pdot}
\end{eqnarray}
Clearly the above equations reduce to Eqn. \ref{eqn:transeqn} in the limit $m\rightarrow 0$ so $\dot{ \mathbf{p}}  \rightarrow 0$). This introduction of artificial inertia is analogous to the widely-used extended  Hamitonian method used in molecular dynamics simulations \cite{Andersen1980, Nose1984, Hoover1985, Car1985}. Yin and Ao provided detailed analysis of this limit, and obtained the corresponding Fokker-Planck equation  \cite{Yin2006},
\begin{eqnarray}
\partial_t \rho = \nabla\cdot \mathbf{M}^{-1}[\epsilon \nabla+\nabla\phi ]\rho
\end{eqnarray}
which can be further rewritten as,
\begin{eqnarray}
\partial_t\rho &=& \nabla\cdot\left[ \epsilon \mathbf{D\cdot\nabla + \epsilon (\nabla \cdot\mathbf{Q}^T) +(\mathbf{D}+\mathbf{Q})\cdot\nabla\phi}  \right]\rho  \label{eqn:FP_yin}
\end{eqnarray}
To derive the above expression we have used the antisymmetric property of $\mathbf{Q}$, and noticed,
\begin{eqnarray}
\nabla\cdot(\mathbf{Q}\cdot \nabla\rho) &=& (\nabla\cdot\mathbf{Q})\cdot(\nabla\rho) = \sum_{ij}\partial_j\left( \frac{\partial Q_{ij}}{\partial x_i} \rho \right) \nonumber\\
&=&-\sum_{ij}\partial_j\left(\frac{\partial Q_{ji}}{\partial x_i}\rho \right) =  \nabla\cdot (\nabla\cdot\mathbf{Q}^T)
\end{eqnarray}

Further studies are needed to examine the general existence and mathematical properties of the transformation. In the next sections we will analyze the validity and implications of this transformation starting from the Fokker-Planck equation corresponding to the SDEs.

\subsection{Flux decomposition}
In this work we will focus on systems for which steady state distributions exist. The following discussions follow the procedure used by Graham and by Eyink {\it et al} closely \cite{Graham1977,Eyink1996}, and provide explicit formula alternative to Eqn. \ref{eqn:M_eqn} for constructing $\mathbf{Q}$. Let's consider a generic Fokker-Planck equation,
\begin{eqnarray}
\partial_t\rho(\mathbf{x},t) &=& \nabla\cdot[-\mathcal{J}\rho(\mathbf{x},t) + \epsilon\mathbf{D}\cdot \nabla \rho(\mathbf{x},t)] \label{eqn:FP_generic}
\end{eqnarray}
Compared to Eqn. 4.1 of Eyink {\it et al.} \cite{Eyink1996}, we use a slightly different form by proper redefinition of  $\mathcal{J}$. As will be clear in the following development, with this redefinition one avoids the requirement that the matrix $\mathbf{D}$ has to be divergence free as Eyink {\it et al.} assumed.  In the following derivations we will not specify the form of $\mathcal{J}$, thus the results are independent of the choice of interpretations of the SDEs, a subtle issue we will address in the next section. This is another aspect in generalizing  the procedure of  Eyink {\it et al.} \cite{Eyink1996}, where the Ito interpretation is adopted.
One can define an "entropy" term from the steady state distribution $\mathcal{S} \equiv \ln(\rho_{ss})$. Notice that one can decompose the flux term $\mathcal{J}$ in Eqn. \ref{eqn:FP_generic} into a conservative part $\mathcal{R}$ and a dissipative part $\mathcal{D}$,
\begin{eqnarray}
\mathcal{D} \equiv  \epsilon\mathbf{D}\cdot\nabla\mathcal{S} =- \mathbf{L}^s\cdot\mathbf{X} , \mathcal{R} \equiv  \mathcal{J}-\mathcal{D} 
\end{eqnarray}
In the above expression we have written in the Onsager form with $\mathbf{L}^s \equiv \epsilon \mathbf{D}$ being a symmetric matrix, and $\mathbf{X} \equiv -\nabla\mathcal{S}$ being the general force. (In this work, we won't distinguish covariant and contravariant tensors in our notations).
Substituting the expression of $\rho_{ss}$ into Eqn. \ref{eqn:FP_generic}, one has,
\begin{eqnarray}
\nabla \cdot \left[ \mathcal{R} \exp(\mathcal{S})  \right] = 0
\end{eqnarray}
Then following Graham, and Eyink {\it et al}, one can relate the divergence free vector $\mathcal{R} \exp(\mathcal{S}) $ to an antisymmetric matrix $\mathcal{F}$,
\begin{eqnarray}
\mathcal{R}_i \exp(\mathcal{S}) \equiv  \sum_j\frac{\partial \mathcal{F}_{ij}}{\partial x_j} \label{eqn:F_def}
\end{eqnarray}
The antisymmetric matrix $\mathcal{F}$ can be further expressed in terms of a vector potential (see below).
So,
\begin{eqnarray}
\mathcal{R}_i = L_{ij}^a\cdot \frac{\partial \mathcal{S}}{\partial x_j} +\frac{\partial L^a_{ij}}{\partial x_j} \label{eqn:R_decomposition}
\end{eqnarray}
with the new antisymmetric matrix $L^a_{ij}\equiv \mathcal{F}_{ij}\exp(-\mathcal{S})$. Therefore the drift term $\mathcal{J}$ can be expressed as,
\begin{eqnarray}
\mathcal{J}_i = -(L^s_{ij}+L^a_{ij})X_j + \sum_j\frac{\partial L^a_{ij}}{\partial x_j} 
\label{eqn:J_decomposition}
\end{eqnarray}
We essentially reproduced the derivation of Eyink {\it et al} but without the need for the divergence free approximation on $\mathbf{L}^s$ and $\mathbf{L}^a$. Notice that in the above derivations no assumption is made on whether the system is linear.
 
Graham discusses how one can construct the matrix $\mathcal{F}$ and so $\mathbf{L}^a$  from the steady state distribution and current with additional constraints (e.g., satisfying homegeneous Maxwell' equation) \cite{Graham1977}. 
The antisymmetric matrix $\mathbf{L}^a$ can only be determined up to a gauge transformation. This can be seen by noticing that ${\mathbf{L}^a}' = \mathbf{L}^a + \chi \exp(\mathcal{-S})$ also satisfies Eqn. \ref{eqn:R_decomposition} provided $\sum_j \nabla_j\chi_{ij}=0$.  Zia and Schmittmann  derived the explicit form of the anti-symmetric matrix for linear Langevin systems \cite{Zia2007}.  We suggest that the decomposition can be viewed as generalization of Helmholtz's theorem, or the fundamental theorem of vector calculus, to arbitrary dimensional systems. 

It is well known that the calculus of a SDE is ambiguous depending on the choice of interpretation. Consequently, for a given Fokker-Planck equation, one can construct different SDEs. While the decomposition of Eqn. \ref{eqn:J_decomposition} is formal, one can use different definitions of $\mathcal{J}$. The definition of $\mathcal{J}$  due to  Ito and Stratonovich, respectively, 
\begin{eqnarray}
\mathcal{J}_{Ito} &=&  \mathbf{G}  -\nabla\cdot\mathbf{D} \nonumber\\
\mathcal{J}_{Stratonovich} &=& \mathbf{G}  
\end{eqnarray} 
Where $\mathbf{G}$ is as defined earlier in Eqn. \ref{eqn:stochaseqn}. One may easily define the corresponding expressions for other proposed interpretations ({\it e.g.} \cite{Hanggi1978, Klimontovich1994, Wong1965}).  For a special choice $\mathcal{J}_{Ao} = \mathbf{G} -   \nabla\cdot (\mathbf{L}^a)^T$, Eqn. \ref{eqn:J_decomposition} gives $ \mathbf{G}=-\mathbf{(L^s+L^a)}\cdot \mathbf{X} $.  The subscript of $\mathcal{J}$ indicates that this choice corresponds to the zero-mass limit interpretation of Ao \cite{AoJPhys2004}. For a system with steady state, Yin and Ao (also see discussions below) showed that the steady state distribution is given by $\rho_{ss}(\mathbf{x})\propto \exp(- \phi(\mathbf{x})/\epsilon)$ \cite{Yin2006}.
In the literature this relation is also widely used to define the potential \cite{Graham1971}. Substituting this expression into Eqn. \ref{eqn:FP_yin}, and comparing the result with Eqn. \ref{eqn:J_decomposition} term by term, one can easily identify that $\mathcal{S} \equiv -\phi/\epsilon$, $\mathbf{L}^s \equiv \epsilon \mathbf{D}$, and $\mathbf{L}^a \equiv  \epsilon \mathbf{Q}$. 

Since $(\mathbf{D}+\mathbf{Q})\cdot \nabla\phi = -\mathbf{G}$ and $\mathbf{Q}$ has an undetermined choice of gauge, a class of SDEs under the zero-mass interpretation can produce a given set of steady state distribution and current.  A given form of $\mathbf{G}$ corresponds to a specific choice of the gauge. As it will be apparent in the discussions below, we suggest that this freedom of gauge choice is mathematically analogous to the gauge freedom in electrodynamics. It is unclear whether  the relations given in Eqn. \ref{eqn:M_eqn} are sufficient to determine $\mathbf{Q}$ uniquely except for simple cases discussed by Kwon {\it et al.} \cite{Kwon2005}.

\subsection{Dependence and independence on Interpretation}
. While Ao has developed the transformation following an unconventional procedure, the essence is that one may define a new interpretation of the SDEs, which has certain appealing mathematical properties as we will discuss below. 
Given the SDEs under one interpretation, there are standard procedures to obtain the corresponding equations under another interpretation \cite{Oksendal2003}. Here we suggest a procedure to construct the SDEs of the zero-mass limit corresponding to {\it e.g.} those of Ito interpretation. First construct the Fokker-Planck equation and the decomposition \ref{eqn:J_decomposition} of the latter, then the SDEs of the zero-mass interpretation are given by the diffusion matrix $\mathbf{D}$ and $\mathbf{G} = -( \mathbf{D+Q})\nabla\phi$. We will show an example below.

In quantum mechanics, one may choose different representations. Analogously one may interpret a given Fokker-Planck equation with different stochastic differential equations. This representation ambiguity is related to the gauge invariance discussed above. In one aspect, a class of stochastic processes may share the same stationary and at least some of the dynamic properties. in another aspect, for a given stochastic process one may use different forms of mathematical description depending on how one abstracts the noise source properties.  However, one expects that the physically measurable quantities should be independent of the choice of interpretation, {\it e.g.},  the values of ensemble averaged quantities since the underlying Fokker-Planck equation is the same. This observation has practical importance. While the mapping discussed in this work allows straightforward theoretical analysis based on knowledges of Hamiltonian systems,  it is still unclear how one can adopt the zero-mass interpretation in numerical simulations, and apply the theoretical results. Instead one can perform the usual stochastic simulations under the Ito or Stratonovich interpretation with the corresponding SDEs.   

\subsection{Mapping}
Now let's define a Langrangian,
\begin{eqnarray}
\mathcal{L}_0&=& \frac{1}{2}m\dot{\mathbf{x}}^2-\phi+\dot{\mathbf{x}}\cdot \mathbf{A}(\mathbf{x})\nonumber
\end{eqnarray}
The Euler-Lagrange equation from variation of $\mathcal{L}_0$, $\delta \mathcal{L}_0 =0$, 
\begin{eqnarray}
\frac{d}{dt} \left(\frac{\partial \mathcal{L}_0}{\partial \dot{x_i}}\right) -\frac{\partial \mathcal{L}_0}{\partial x_i} =0 \nonumber
\end{eqnarray}
results in Eqns. \ref{eqn:xdot} and \ref{eqn:pdot} without the dissipative terms in Eqn. \ref{eqn:pdot}, provided,
\begin{eqnarray}
T_{ij} = \frac{\partial A_i}{\partial x_j} - \frac{\partial A_j}{\partial x_i} \nonumber
\end{eqnarray}
Therefore, $\mathbf{T}$ is analogous to the electromagnetic tensor \cite{Jackson1999}.  One can also define the conjugate momentum, 
\begin{eqnarray}
\tilde{p}_i = \frac{\partial \mathcal{L}_0}{\partial \dot{x_i}} = p_i + A_i(\mathbf{x}) \nonumber
\end{eqnarray}
and a corresponding Hamiltonian,
\begin{eqnarray}
\mathcal{H}_0 = \mathbf{\tilde{p}\cdot \dot{x}} -\mathcal{L}_0 = \frac{(\mathbf{\tilde{p}-A(x)})^2}{2m}+\phi(\mathbf{x})
\end{eqnarray}

Next following Zwanzig \cite{Zwanzig1973}, we will show that the dissipative terms in Eqn. \ref{eqn:pdot}  can be replaced by a harmonic bath hamiltonian.
First, let's summarize the result of Zwanzig. Consider a system described by a set of state variables 
$\mathbf{z}\equiv (\mathbf{x, p})$ (for coordinates and corresponding momenta), 
and similar bath variables $\mathbf{Y}$. Assume that one can define a function
 \begin{eqnarray}
 &&\mathcal{H}(\mathbf{z,Y})= \mathcal{H}_s(\mathbf{z})+\mathcal{H}_b(\mathbf{z,Y}) \nonumber\\
&&=  \mathcal{H}_s(\mathbf{z}) + \frac{1}{2} \mathbf{[ Y-a(z)]}^T \cdot \mathbf{K}\cdot [ \mathbf{Y-a(z)}]
 \end{eqnarray} 
 so
\begin{eqnarray}
d\mathbf{z}/dt &=& \mathbf{J\cdot \nabla_z}(\mathcal{H}_s+\mathcal{H}_b) \nonumber\\\
d\mathbf{Y}/dt &=& \mathbf{J\cdot \nabla_Y} \mathcal{H}_b \nonumber
\end{eqnarray}
with
\begin{eqnarray}
\mathbf{J}= \left( \begin{array}{cc} 
0 & I\\
-I & 0
\end{array}
\right) \nonumber
\end{eqnarray}
where $I$ is the identity matrix having the dimension of the total coordinates (or momenta). Then one can integrate out the bath variables, and obtain (assuming that the initial bath variables $\mathbf{Y}_0$ are drawn at random from a canonical distribution, {\it i. e.}, in contact with a heat bath),
\begin{eqnarray}
&&d\mathbf{z}_t/dt = \mathbf{J\cdot \nabla_z} \mathcal{H}_s(\mathbf{z}_t) +\mathbf{J \cdot W(z}_t)\cdot \mathbf{F}(t) \nonumber\\
&&+ \int_0^t d\tau \mathbf{J \cdot W(z}_t)\cdot \mathbf{\Gamma}(\tau)\cdot \mathbf{W}^T (\mathbf{z}_{t-\tau})\cdot  \dot{\mathbf{z}}_{t-\tau} 
\label{eqn:GLE}
\end{eqnarray}
where,
\begin{eqnarray}
\mathbf{W(z}) &=& \mathbf{\nabla_{z}a}^T(\mathbf{z}) \nonumber\\
\mathbf{F}(t) &=& -\mathbf{K} \cdot \exp(t \mathbf{J\cdot K})\cdot [\mathbf{Y_0-a(z_0)}] \nonumber \\
<\mathbf{F}(t)> &=& 0 \nonumber \\
<\mathbf{F}(t)\mathbf{F}^T(\tau)> &=& \epsilon \mathbf{\Gamma} (t-\tau) \nonumber \\
\mathbf{\Gamma}(t) &=& \mathbf{K}\cdot \exp(t \mathbf {J\cdot K}) \nonumber
\end{eqnarray}
For our purpose let's choose,
\begin{eqnarray}
\mathcal{H}_b = \sum_{\alpha=1}^{N_\alpha} 
         \sum_{j=1}^N \left( \frac{1}{2} p_{\alpha j}^2 
         + \frac{1}{2}\omega_{\alpha j}^2(q_{\alpha j}-a_{\alpha}(\mathbf{x})/(\sqrt{N} \omega_{\alpha j}^2))^2 
         \right) 
\end{eqnarray}
Then Eqn. \ref{eqn:GLE} becomes,
\begin{eqnarray}
d\mathbf{x}_t/dt &=& \nabla_{\mathbf{p}} \mathcal{H}_s \\
d\mathbf{p}_t/dt &=& - \nabla_{\mathbf{x}} \mathcal{H}_s  \nonumber\\
&+& \sum_{\alpha=1}^{N_\alpha}  \left[ - \int_0^t d\tau \gamma_\alpha(\tau) \mathbf{p}_{t-\tau}/M
+\mathbf{F}_\alpha(t) \right]  \label{eqn:GLE_system}
\end{eqnarray}
where,
\begin{eqnarray}
\gamma_\alpha(t) &=& \frac{1}{N}(\mathbf{\nabla_{x}}a_\alpha)^T (\mathbf{\nabla_{x}}a) \sum_{j=1}^{N} \frac{1}{\omega_j^2} \cos(\omega_j t) \nonumber\\
\mathbf{F}_\alpha (t) &=& (\mathbf{\nabla_{x}}a_\alpha)   \sum_{j=1}^N  \left[  \left(q_{\alpha j}(0)-\frac{a_\alpha(\mathbf{x}_0)}{\sqrt{N} \omega_{\alpha j}^2} \right) \cos(\omega_{\alpha j} t) \right. \nonumber\\
&& \left. + p_{\alpha j}(0) \frac{\sin(\omega_{\alpha j}t)}{\omega_{\alpha j}} \right] \nonumber
\end{eqnarray}
Let's use  a frequency distribution
\begin{eqnarray}
\rho(\omega) =  \left\{ \begin{array}{ll} 
3 \omega^2/\omega_d^2, & \omega<\omega_d \nonumber\\
0, & \omega>\omega_d 
\end{array} \right.
\end{eqnarray}
where $\omega_d$ is a cutoff frequency.  We replace the sums over the oscillator frequencies by integrals,
\begin{eqnarray}
\sum_{j=1}^N \rightarrow N\int_0^{\omega_d} d\omega \rho(\omega), \nonumber
\end{eqnarray}
and assume that the system momenta do not change significantly within the time interval $1/\omega_d$. Then Eqn. \ref{eqn:GLE_system} can be approximated as,
\begin{eqnarray}
d\mathbf{p}_t/dt &=& - \nabla_{\mathbf{x}} \mathcal{H}_s  + \sum_{\alpha=1}^{N_\alpha}  \left[ - \gamma_0 (\mathbf{\nabla_{x}}a_\alpha^T) (\mathbf{\nabla_{x}}a_\alpha^T)^T  \mathbf{p}_{t}/M \right.  \nonumber\\
&& \left. +\sqrt{2 \gamma_0\epsilon }  (\mathbf{\nabla_{x}}a_\alpha)  \zeta_\alpha(t) \right]
\end{eqnarray}
where $\gamma_0 = 3\pi/(2\omega_d^2)$, and $\zeta_\alpha$ is a random number drawn from a normal Gaussian white noise distribution.

Consider the dissipative terms $- \mathbf{S}\frac{d\mathbf{x}}{dt} +\mathbf{g'(x)}\zeta(t) $ in the special case that  $\mathbf{G = L\cdot x}$ with $\mathbf{L}$ a constant matrix, and the diffusion matrix $\mathbf{D}$
can be approximated as a constant matrix. Then $\mathbf{S}$ and $\mathbf{g'(x)}$ are constants \cite{Kwon2005}. 
This corresponds to a linear expansion of 
a system around a stable stationary point of the steady state distribution. In this case the dissipative terms can be replaced by a bath Hamiltonian,
\begin{eqnarray}
\mathcal{H}_b = \sum_{\alpha=1}^{m} \left[ 
         \sum_{j=1}^{N} \left( \frac{1}{2} p_{\alpha j}^2 
         + \frac{1}{2}\omega_{\alpha j}^2 \left(q_{\alpha j}-    \frac{\sum_{i=1}^{n}  g'_{i \alpha } x_i} {\sqrt{ 2N  \gamma_{ 0}\epsilon}  \omega_{\alpha j}^2} \right)^2 
         \right) \right]  \label{eqn:Linearbathhamiltonian}
\end{eqnarray}
where $n\times m$ is the dimension of $\mathbf{g'}$. The term $\nabla \mathbf{a}$ can be geometrically interpreted as $n$ vectors $\{\mathbf{v}_1,\cdots, \mathbf{v}_n \}$ in the $m$ dimensional Euclid space, and 
$S_{ij} \propto \mathbf{v}_i\cdot \mathbf{v}_j = \vert   \mathbf{v}_i \vert   \vert   \mathbf{v}_j \vert  \cos(\theta_{ij})$. 

To prove existence of the bath Hamiltonian in the general case, one needs to show that for any given diffusion matrix $\mathbf{S= g'g'}^T/2$, there is at least one set of  $\mathbf{a}$ 
so $\mathbf{S} =\nabla \mathbf{a}^T \cdot (\nabla \mathbf{a}^T )^T$.  Notice that the needed number of $a_\alpha$, $N_\alpha$, can be any positive integer. Let's assume that  $g'_{ij}$ are at least piecewise analytical, and can be expressed as s-th order polynomial around one point $\mathbf{x}_0$,
\begin{eqnarray}
g'_{ij} = g'_{ij}(\mathbf{x}_0) + \nabla g'_{ij}(\mathbf{x}_0) \cdot \Delta \mathbf {x} + \cdots \nonumber
\end{eqnarray}
Then one can choose 
\begin{eqnarray}
a^{\alpha} =  \mathbf{a}_1^{\alpha}(\mathbf{x}_0) \cdot \Delta \mathbf {x} + \frac{1}{2} \Delta \mathbf {x}^T  \cdot  \mathbf{a}_2^{\alpha}(\mathbf{x}_0) \cdot \Delta \mathbf {x}  +  \cdots \nonumber
\end{eqnarray}
where $\mathbf{a}_1^\alpha$ are vectors, and $\mathbf{a}_n^\alpha$ are $n$-th ranked  symmetric  tensors. Then,
\begin{eqnarray}
 S_{ij} &=& \frac{1}{2} \sum_k  \left[ g'_{ik}  g'_{jk}  + \left( g'_{jk} \nabla g'_{ik} \right. \right. 
     \left.  \left. + g'_{ik} \nabla g'_{jk} 
         \right) \cdot \Delta \mathbf {x}  
      \right] \nonumber\\
     && +   \cdots \label{eqn:Dexpansion}
 \end{eqnarray}
 \begin{eqnarray}
 &&(\nabla a^T (\nabla a^T)^T)_{ij} =   \sum_\alpha  \left[ (a^{\alpha}_1)_i (a^{\alpha}_1)_j  \right. \nonumber\\
 &&\left. +  \sum_k \left( (a_1^\alpha)_j   (a_{2}^\alpha)_{ik}   +  
               (a_1^\alpha)_i   (a_2^\alpha)_{jk}  
              \right) \Delta {x}_k 
 \right]      + \cdots \label{eqn:Aexpansion}
\end{eqnarray}
Let's equate the above two expressions term by term. For the terms in the $i$-th power of $\Delta \mathbf{x}$, Eqns. \ref{eqn:Dexpansion} and \ref{eqn:Aexpansion} give $n (n+1)/2 \times (i+n-1)!/((n-1)!i!)$ relations, which are linear equations for the elements of $\mathbf{a}_{i+1}^\alpha$.  Exceptions are for the zero-th order terms, which give quadratic equations.  There are  $N_\alpha \times (i+n)!/((n-1)!(i+1)!)$ elements of  $\mathbf{a}_{i+1}^\alpha$ need to be determined. Therefore the number of variables is no less than the number of constraints provided $N_\alpha\geq n(n+1)(s+1)/(2s+2n)$. For the zero-th term, one can always rotate the vectors described 
in Eqn. \ref{eqn:Linearbathhamiltonian} in the expanded $N_\alpha$ dimensional space, which is still an acceptable solution. Especially one can construct a solution  so that each vector has nonzero projection on each axis. Then the higher order terms $\mathbf{a}_i^\alpha$  with $i>1$ can be determined (if the number of variables equals the number of constraints) or selected (if the number of variables is larger than the number of constraints) in sequence.  Consequently, in practice one can always relate $\mathbf{S}$ to a set of scalar functions $\{a^\alpha\}$ to any necessary accuracy.

Therefore one can map a dissipative system described by Eqns \ref{eqn:stochaseqn}  to a thermostated Hamiltonian system with $\mathcal{H}=\lim_{ m\rightarrow 0,\omega_b\rightarrow\infty}(\mathcal{H}_0+\mathcal{H}_b)$. Caution should be taken since that the two limits may not commute. This concludes our proof.
The Hamiltonian mathematically corresponds to Dirac's constrained Hamiltonian \cite{Dirac2001}. It describes a massless particle, coupled to a set of harmonic oscillators, moving in a hypothetical scalar potential and vector (magnetic) potential field expanded in $n$-dimensions. The magnetic field arises from breaking of detailed balance.
The steady state distribution is thus given by the Boltzmann distribution of the Hamiltonian system,
\begin{eqnarray}
\bar{\rho}_{ss} (\mathbf{x,p,Y}) = \frac { \exp(- \mathcal{H}/\epsilon )}
 {\int d\mathbf{x}d\mathbf{p} 
d\mathbf{Y} \exp(- \mathcal{H}/\epsilon )}\label{eqn:rho_ss1}
\end{eqnarray}
or after integrating the momenta, and the bath variables,
\begin{eqnarray}
\rho_{ss}(\mathbf{x}) = \frac { \int d\mathbf{p} d\mathbf{Y}  \exp(-\mathcal{H} /\epsilon)}
 { \int d\mathbf{x}d\mathbf{p} 
d\mathbf{Y}\exp(- \mathcal{H} /\epsilon)} \propto \exp(- \phi/\epsilon) \label{eqn:rho_ss2}
\end{eqnarray}
where $\mathbf{Y}$ represents all the bath variables. Eqn. \ref{eqn:rho_ss2}  is conjectured by Ao \cite{AoJPhys2004}, and is consistent with what has been proposed by others\cite{Oono1998,Graham1971}. Here we reach the same conclusion naturally, and provide the theoretical basis for further studies. It may at first seem surprising that a nonequilibrium steady state with nonzero currents can be mapped to an equilibrium state. The zero mass limit is the key. This may share interesting analogies with superconductivity.  

\section{Examples}
To further clarify the above abstract discussions, let us consider a simple analytically solvable example,
\begin{eqnarray}
\frac{dx}{dt} = 1-x + \sqrt{2} \zeta_1(t), \frac{dy}{dt} = x - y +\sqrt{2}\zeta_2(t)
\end{eqnarray}
with the diffusion constant matrix set to identity. The matrix $\mathbf{Q}$ can be calculated from Eqn. \ref{eqn:M_eqn},
\begin{eqnarray}
\mathbf{Q} = \frac{1}{2} \left(\begin{array}{cc} 0 &1\\-1 & 0 \end{array}\right)
\end{eqnarray}
Consequently,
\begin{eqnarray}
&&-\nabla\phi = (\mathbf{D}+\mathbf{Q})^{-1}\cdot \mathbf{G} =\frac{2}{5} (-3x+y+2,x-2y+1)^T \nonumber\\
&& \phi = \frac{1}{5}\left( 2(x-1)^2+(x-y)^2+(y-1)^2\right) +\phi_0
\end{eqnarray}
The gauge invariance suggests that the same steady state distribution and current are given by the following stochastic system (under the zero-mass  interpretation),
\begin{eqnarray}
\frac{d}{dt}\left(\begin{array}{c}x\\y \end{array} \right) = \left(\begin{array}{c} 1-x \\ x-y  \end{array} \right) - \Delta\mathbf{Q}\cdot\nabla\phi + \sqrt{2}\left(\begin{array}{c}\zeta_1(t)\\\zeta_2(t) \end{array} \right)
\end{eqnarray}
where
\begin{eqnarray}
\Delta \mathbf{Q} =  \left(\begin{array}{cc} 0 & c\\-c &0 \end{array}\right) \exp(\phi)
\end{eqnarray}
with $c$  an arbitrary  constant. On the other hand,one cannot make the same statement  in general under the Ito or Stratonovich interpretation. In their analysis based on master equations, Zia and Schmittmann  also have a discussion highlighting the fact that  more than one dynamic system can lead to the same steady state distribution and current \cite{Zia2007}.

As another example, let's consider a system adopted by Gomez-Solano {\it et al.} to test a modified fluctuation-dissipation theorem \cite{Gomez-Solano2009}. In the experimental setup, a bead is confined to
a circle, which has a radius as $r_e=1$ with proper choice of units). The bead experiences a potential $\sin\theta$ and a constant force $F$. The corresponding Langevin equation is  
\begin{eqnarray}
d\theta /dt= -H_0\cos\theta + F+\sqrt{2D} \zeta_\theta(t) \label{eqn:dtheta}
\end{eqnarray}
with $<\zeta_\theta(t)\zeta_\theta(t')>=\delta(t-t')$, and here for simplicity we take $\epsilon=k_BT=1$.  In the original work the Ito or Stratonovich interpretation is adopted. The motion along the radial direction is highly confined. The model resembles closely to rotary molecular motors, where $F$ can be the applied external torque or the nonequilibrium driving force
from transmembrane electrochemical potential or chemical energy ({\it e.g.}, ATP hydrolysis). Mathematically the model represents an even larger class of systems under external force and periodic potential, such as linear translational molecular motors like kinesin \cite{Reimann2002}.
The "steady state" distribution (projected to and normalized within a range of $2\pi$) is \cite{Reimann2002},
\begin{eqnarray}
\rho_{ss}^\theta(\theta) = \mathcal{N} \int_0^{2\pi}d\theta' e^{\left[\left( H_0(\sin(\theta+\theta')-\sin\theta)-F\theta'\right)/D\right]}
\end{eqnarray}
where $\mathcal{N}$ is the normalization factor. The quotation mark on "steady state" emphasizes that rigorously speaking it is not a steady state in the actual range $\theta=(-\infty,\infty)$.  Consequently it is incorrect to directly map the nonequilibrium steady state  to the
equilibrium state of a  Hamiltonian system with a potential $(H_0\sin\theta-F\theta)$.
Instead let us rewrite Eqn. \ref{eqn:dtheta} and include  an additional equation for the radial coordinate,
\begin{eqnarray}
d\theta /dt &=& -\frac{H_0}{r^2}\cos\theta + \frac{F}{r^2}+\sqrt{2D/r^2} \zeta_\theta(t) \nonumber\\
dr/dt  &=&  -\kappa (r-1) +\frac{D}{r}+\sqrt{2D} \zeta_r (t)
\end{eqnarray}
with $<\zeta_r(t)\zeta_r(t')>= \delta(t-t')$, $<\zeta_r(t) \zeta_\theta(t')>=0$, and $\kappa>>1$. Transform the Langevin equations into  Cartesian coordinates,
\begin{eqnarray}
\frac{dx}{dt} = -[\kappa (r-1) - \frac{D}{r}] \frac{x}{r}  + H_0\frac{xy}{r^3} -F\frac{y}{r^2} + \frac{\sqrt{2D}}{r} ( x\zeta_r - y\zeta_\theta) \nonumber\\
\frac{dy}{dt} =  -[\kappa (r-1) - \frac{D}{r}] \frac{y}{r}  - H_0\frac{x^2}{r^3} +F\frac{x}{r^2} + \frac{\sqrt{2D}}{r} ( y\zeta_r + x\zeta_\theta) \nonumber
\end{eqnarray}
The diffusion matrix and the flux in the Cartesian coordinate is,
\begin{eqnarray}
\mathbf{D} &=& 
\left(\begin{array}{cc}
D &0\\
0&D\end{array}\right)
\nonumber\\\\
\mathcal{J} &=& \left(\begin{array}{c}
-[\kappa (r-1) - \frac{D}{r}] \frac{x}{r}  + H_0\frac{xy}{r^3} -F\frac{y}{r^2} \\
 -[\kappa (r-1) - \frac{D}{r}] \frac{y}{r}  - H_0\frac{x^2}{r^3} +F\frac{x}{r^2} 
\end{array}\right)  \nonumber\\
&&- \nabla\cdot\mathbf{D}
\end{eqnarray}
For this system $\nabla\cdot\mathbf{D}=0$.
One can show that the overall steady state distribution is then given by,
\begin{eqnarray}
\rho_{ss}(r,\theta) \propto r \rho_{ss}^\theta(\theta)\exp\left[-\frac{1}{2D} \kappa(r-1)^2 \right]
\end{eqnarray}
Then from Eqn. \ref{eqn:F_def}, 
\begin{eqnarray}
&&Q_{12}(x,y) = \frac{\rho_{ss}(r_0,\theta_0)}{\rho_{ss}(r,\theta)} Q_{12}(x_0,y_0)   \nonumber\\
&& + \frac{1}{\rho_{ss}(r,\theta)} \int_{(x_0,y_0)}^{(x,y)}\rho_{ss}(r',\theta') (dy,-dx) \cdot \nonumber\\
&& [\mathcal{J}(x',y') - (\mathbf{D}\cdot  \nabla\ln\rho_{ss}(r',\theta')] \label{Eqn:Q12}
\end{eqnarray}
where the integration path connects a reference point $(x_0,y_0)$ and the end point $(x,y)$, and $Q_{12}(x_0,y_0)$ determined with Eqn. \ref{eqn:J_decomposition} at the reference point.
Let us focus on the case where the motion along the radial direction is confined to $r=1 (\kappa>>1)$. 
Then with the integration path along the unit circle Eqns. \ref{Eqn:Q12} reduces to,
\begin{eqnarray}
Q_{12}(\cos\theta,\sin\theta)  
\propto \frac{1}{\rho^\theta_{ss}(\theta)}
 \label{Eqn:Q12_reduced} 
%
\end{eqnarray} 
One can then construct the corresponding SDEs under the zero-mass interpretation. Therefore, one can equivalate the original nonequilibrium dynamic system to a massless particle moving along a 1-D circular track with potential $\phi= -\ln\rho_{ss}^\theta(\theta)$, and a nonuniform magnetic field perpendicular to the plane. The magnetic field can be determined from the antisymmetric part of $(\mathbf{D+Q})^{-1}$. This result can be tested experimentally. 
 
\section{Discussions and conclusions}
For systems having a steady state, we demonstrate the relation between the transformation introduced by Ao and the decomposition proposed by Graham and by Eyink et al. We suggest that the decomposition is an application of the  generalized Helmholtz  theorem in vector field analysis to arbitrary dimensions. The theorem states that any vector field in three dimensions satisfying certain mild mathematical conditions can be decomposed into a divergence-free portion and a rotation-free portion. We further prove that one can map the stochastic and generally non-Hamiltonian system into a Hamiltonian system.  The mapping allows one to study dynamics of a general system with the techniques developed for Hamiltonian systems, especially for equilibrium Hamiltonian systems. If one can prove the generalized Hemholtz's theorem, which may be related to the Hodge decomposition \cite{Bott1982}, then we suggest that the results in this work may hold for systems without steady state. 
The results may even be generalized to discrete dynamics \cite{Zia2007, Jiang2003}. Further studies are necessary to clarify the mathematical structure and to develop numerical algorithms. We hope that this work may inspire further dialogue between researchers in nonequilibrium statistical physics (especially dynamics of chemical networks) and in other fields such as plasma, superconductivity, quantum field theory, and liquid crystal physics.

In real applications, constructing the exact form of the overall mapping Hamiltonian is impractical.  
Most existing theories on nonequilibrium steady states require knowledge of the steady state distributions and currents \cite{Graham1977,Eyink1996,Hanggi1990,   Zia2007}. The procedure of constructing the transformation matrices given in this work is of no exception to this requirement.  Due to violation of the detailed balance in general there is no readily found stationary distribution as for the equilibrium systems. In principle Eqn. \ref{eqn:M_eqn} leads to the Boltzmann-like distribution. However solving Eqn. \ref{eqn:M_eqn} may be even more difficult than solving the corresponding Fokker-Planck equation directly. The transformation leads to a new zero-mass interpretation of the SDEs. It is unclear how to develop the corresponding numerical algorithms. 

The question is then why take the  effort of introducing this transformation and new interpretation? We want to emphasize that the major result of this work is the existence of the mapping itself. Hamiltonian systems have special mathematical structures and properties. As stated by Morrison \cite{Morrison1986,Morrison1998},  the mapping to a Hamiltonian system allows one to utilize the special properties of the Hamiltonian system to guide further understanding of  general dynamical systems, which may  not been so transparent otherwise. 
Unlike equilibrium statistical physics, the lack of  a general framework seriously impedes the
 development of nonequilibrium theories. The current result provides such a framework. 
The mapping allows one to define a partition function, and thus some pseudo-thermodynamic quantities for a dissipative system. Then one can readily apply well-developed equilibrium results to nonequilibrum systems. The last one or two decades have witnessed important progress in nonequilibrium physics studies, and several relations concerning the nonequilibrium steady states have been derived. With the mapping,  we now know that these results, previously derived through some special techniques and from phenomenological frameworks, must exist, simply because there are counterparts for the equilibrium states.
Let's just mention a few examples. The result of Hatano and Sasa (Eqn. 11 of \cite{Hatano2001}) turns out to be  the equilibrium Jarzynski equality \cite{Jarzynski1997}.  One can develop theories on noise-induced phase transitions \cite{Horsthemke1984, Broeck1994} parallel to those of thermodynamic phase transitions, noticing that for the former case the noise amplitude replaces the temperature (refer to discussions below Eqn. \ref{eqn:stochaseqn}). The fluctuation-dissipation relations are originally derived for processes near equilibrium \cite{Nyquist1928, Callen1951,Stratonovich1993}. Recent years there are many efforts on generalizing it to systems far from equilibrium 
\cite{Prost2009, Graham1977, Eyink1996,Cugliandolo1994,Marconi2008}. The mapping between a nonequilibrium steady state and a thermodynamic equilibrium state makes the generalization straightforward \cite{Graham1977, Eyink1996, XingFD2009}. The existence of the mapping also allows us to derive a  Zwanzig-Mori projection formula for general non-Hamiltonian dynamics \cite{XingMZ2009}, where defining an invariant measure is a major obstacle in earlier attempts \cite{Chorin2000,Chorin2002}. Biological networks are important nonequilibrium systems.  They are typically highly inhomogeneous, and full of competing interactions. Thus the dynamics of a biological network in some sense resembles that of spin glasses but with nonrandom quenching \cite{Fischer1993}. It is interesting to notice that the mapping between a dissipative and a Hamiltonian system suggest possible glassy behaviors for relaxation to a nonequilibrium steady-state.  

When a slow classical Hamiltonian system couples through positional coordinates to a fast and chaotic classical Hamiltonian system, to a first order approximation the fast system exerts on the slow system velocity-dependent forces described by a symmetric and  an antisymmetric tensors \cite{Berry1993, Rau1997}. The latter is called "geometric magnetism", analogous to the "geometric phase". It is not a conincidence that it resembles the theory discussed here. While the former emphasizes the dynamic effect,  and this (and Ao's) work  emphasizes the thermodynamic origin, the source of magnetism is due to the fact that the slow system is out of equilibrium. For the systems considered here, the fast systems are described by white noises and thus chaotic. Therefore the current formalism, which applies to non-Hamiltonian systems, can be viewed as a more general description of the phenomenon.       

Eqn. \ref{eqn:stochaseqn} contains white noise. For systems subject to colored noise, in principle one can remove them by introducing auxiliary degrees of freedom. The mapping introduced by Ao requires existence of  noise. However, there is no constraint  on the noise strength. Therefore one might use the method even to extrapolate results of deterministic cases.  For systems with spatial dependence, {\it e.g.} the Kardar-Parisi-Zhang equation \cite{Kardar1986}, in practice one can always replace the corresponding stochastic partial differential equations into ordinary stochastic differential equations through grid discretization.  

In this work we derived the mapping between classical non-Hamiltonian and Hamiltonian systems. The quantum counterpart of Eqn. \ref{eqn:stochaseqn} and its relation to the classical Langevin equation has been widely discussed \cite{Feynman1963, Caldeira1983,Caldeira1983b,Hakim1985,Marianer1985,Legget1987,Grabert1988,Cohen1997,Hanggi2005}. It is of theoretical interest to generalize the current work to quantum systems especially in the area of quantum dissipation and decoherence for a system approaching a nonequilibrium steady state.

Ao {\it et al.}  argues that the zero mass limit corresponds to a new interpretation of the stochastic differential equations alternative to the usual interpretations of Ito and of Stratonovich \cite{Kubo1991, Ao2007}. The ambiguity may arise for the following reason: even if the diffusion matrix $\mathbf{D}$ is independent of $\mathbf{x}$, after the  transformation, the corresponding matrix $\mathbf{T}$ may depend on $\mathbf{x}$ nonlinearly. Our mapping to a Hamiltonian system suggests that the zero-mass limit interpretation is physically self-consistent. As expected, the ambiguity disappears if the Fokker-Planck equation is given. 


\section{Acknowledgements.}
We thank Royce Zia, Ping Ao, Lan Yin,  Kenneth S Kim, Zhanghan Wu, Hong Qian, Jin Wang, and Weihua Mu for discussions.
 and Dr. Sergio Ciliberto for pointing to ref. \cite{Gomez-Solano2009}.

\end{document}